\documentclass[prl,twocolumn,amsmath,amssymb]{revtex4}

\usepackage{graphicx}
\usepackage{bm}
\usepackage[usenames,dvipsnames]{color}
\definecolor{altncolor}{rgb}{0,0,0.8}
\usepackage[colorlinks=true, linkcolor=green, anchorcolor=altncolor,
citecolor=altncolor, filecolor=altncolor, menucolor=altncolor,
urlcolor=altncolor]{hyperref}

\begin{document}

\title{A quantum phase transition implementation of quantum measurement}

\author{Peter B. Weichman}

\affiliation{BAE Systems, FAST Labs, 600 District Avenue, Burlington, MA 01803}




\begin{abstract}

A model of quantum measurement, illustrated using the spin--boson model, is formulated in terms of a cascading pair of quantum phase transitions. The first produces the desired superposition of macroscopic responses to the microscopic state under investigation, while the second provides a mechanism for subsequent ``wavefunction collapse,'' suppressing superpositions of distinct macroscopic states, producing instead a density matrix that implements the expected classical observation outcome via the Born probability rule. Motivated by numerous carefully designed measurements that may occur during the course of a quantum computation, effects of entanglement are investigated when the measurement is performed on only a subset of the microscopic degrees of freedom.

\end{abstract}

\maketitle

The effects of a noisy environment on a microscopic quantum degree of freedom (e.g., spin or trapped particle) has long been a topic of intense interest, with applications to nuclear, optical, and condensed matter systems (e.g., Ref.\ \cite{SB2010} and references therein). It has also long been understood that quantum phase transitions in such models can induce classical behavior, e.g., localizing a particle to one side or the other of a symmetric double-well potential through suppression of tunneling \cite{L1987,Mermin1991}. Here, through coupling to the environment, symmetries of the quantum state can be broken by nucleating a macroscopic response---essentially coupling the microscopic variable to an infinite mass shift. The environment response is also be readily detectable, permitting inference of statistical properties of the original quantum degree of freedom through repeated experiments.

Here we extend these ideas to a potentially useful model of quantum measurement, namely a specific mechanism for amplification of quantum state properties into a quantitative classical response, explicitly reproducing the expected Born probability rule---but entirely within the many body Schr{\"o}dinger equation \cite{Zurek2019}. Here, rather than simply observing, e.g., the effects of suppressed tunneling, the objective is to implement the environmental coupling in such a way as to induce a desired macroscopic response that effectively measures the state of the system when it has well defined quantum numbers (e.g., spin definitely either up or down). However, it also produces a definite, but effectively random, result when it does not (e.g., arbitrary superposition of up and down), but with probabilities consistent with the Born rule.

This connection between wavefunction amplitudes and classical probabilities requires a second, higher level, quantum phase transition. Thus, the first transition actually produces a superposition of macroscopic states that preserves the original spin state amplitudes. However, the latter is dynamically unstable to even very weak higher level environmental interactions with the spin macrostate degrees of freedom, e.g., stray electromagnetic (EM) interactions with material matrix distortions. When initiated, this interaction induces a second transition (which might be viewed as an explicit dynamical implementation of ``wavefunction collapse''), randomly in favor of one or the other states in proportion to the Born probability. We analyze in some detail a simple, effective model of this process, derived as a mesoscale approximation to an EM-type microscopic interaction. As for any phase transition (or quantum dissipation problem) despite observed macroscopic irreversiblity, the dynamics remains unitary as the extra information spreads (irretrievably) through the environment.

We focus for simplicity on ground state properties. Generalization to finite $T > 0$ is straightforward, being a matter of time scales. Strictly speaking the symmetry breaking of the first transition (being 0+1 dimensional) is destroyed by arbitrarily weak thermal fluctuations. However, for reasonable parameters, the time scale of the latter will be very long compared to the measurement time.

For simplicity, we also treat the two phase transitions as occurring sequentially (i.e., assuming perfect isolation from the broader environment for the first). In reality the two overlap: as the growing macroscopic superposition induced by the first reaches critical size, the decoherence rate induced by the second increases rapidly and leads to wavefunction collapse---a deeper analysis of this process is left for future work. This randomly selected ``pointer state'' \cite{Zurek2019} ultimately reaches macroscopic size where it can be classically probed. Once selected, the stability of the underlying broken symmetry state ensures that the higher level environment now plays a negligible role. The idea of ``quantum Darwinism,'' i.e., the proliferation of copies of the microstate information \cite{Zurek2019}, is exhibited as well: stability of the final macrostate permits many independent measurements from which one can derive the selected quantum number (spin up or down) with negligible effect on the state or the underlying spin.

Given the above characterization of the measurement process, results of entangled spin measurements are considered. Spins that are physically separated following entanglement are selectively brought into contact with independent environments. By applying the above dynamics separately in each Hilbert space one generates well defined quantum states in which the spins remain entangled, but superpositions of macroscopically distinct states are avoided. Measurement results are shown to follow conventional predictions based on the reduced density matrix for the spin degrees of freedom. Such subsystem measurement sequences are an integral part of quantum computation, especially maintenance of topologically protected qubits \cite{SurfCodes2012}.

The main result here is a controlled measurement model, enabling quantitative study through application of existing exact and numerical results for quantum impurity-type models \cite{SB2010,L1987}. There are many measurement protocols that do not fit within this quasistatic, quasi-equilibrium dynamics paradigm, but are instead intrinsically nonequilibrium. For example, photon detection methods in which the photon is destroyed, replaced, e.g., by a greatly amplified, bias voltage-induced electric current. Properly modeling the quantum-to-classical crossover in such systems require different approaches. Photons, in contrast to compact microscopic spin-like degrees of freedom, are extended objects, being excitations of a quantum field that extends correlations over macroscopic, even continental ranges---the field itself might be viewed as a quantum critical object. Thus, double-slit type experiments, and longer range measurements involving entangled photons, require a different setup, but can still benefit from extensions of the underlying ideas \cite{W19photon}.

\paragraph{Quantum spin and its environment:} The spin--boson model describes a single ``impurity'' interacting with a dissipative environment. The impurity is a spin ${\bf S}$, coupled to an infinite bath of harmonic oscillators characterized by a set of boson operators $\hat b_k,\hat b_k^\dagger$, with standard commutation relations $[\hat b_k, \hat b_{k'}^\dagger] = \delta_{kk'}$. The Hamiltonian is $H_\mathrm{SB} = H_S + H_B + H'_{SB}$ with
\begin{eqnarray}
H_S &=& - h_z  \hat S_z - h_x \hat S_x,\ \
H_B = \sum_k \omega_k \hat b_k^\dagger \hat b_k
\nonumber \\
H'_{SB} &=& -\hat S_z \sum_k \lambda_k (\hat b_k + \hat b_k^\dagger).
\label{1}
\end{eqnarray}
The environmental coupling parameters $\lambda_k$ represent a fluctuating addition to the magnetic field component $h_z$. Their strength is characterized by a spectral function
\begin{equation}
J(\omega) = \pi \sum_k \lambda_k^2 \delta(\omega - \omega_k)
= 2\pi \alpha \omega^s e^{-\omega/\omega_c},
\label{2}
\end{equation}
corresponding to scaling $\pi \lambda_k^2 = dk J(\omega_k) (d\omega_k/dk)$ in the limit of an infinite number of degrees of freedom with continuous index $k$: $\sum_k dk \to \int dk$. The final model form embodies a number of possible physical realizations, parameterized by low frequency exponent $s$, controlled by a combination of system dimensionality $d$ and dispersion relation $\omega({\bf k})$, high frequency cutoff $\omega_c$ (e.g., phonon Debye frequency), and overall strength $\alpha$. The case $s = 1$ corresponds to Ohmic dissipation. An alternative, containing essentially identical physics, is the Caldeira--Leggett model \cite{L1987}, replacing ${\bf S}$ by a particle in a double well potential. Here we treat the unbiased case $h_z = 0$, equivalent to symmetric double well, and begin with an isolated spin-$\frac{1}{2}$, $\alpha = 0$ and $h_x = 0$, in some initial $z$-quantized state $|\psi_0 \rangle = C_+ |+\rangle + C_- |-\rangle$.

\paragraph{Quantum criticality:} The Feynman path integral formalism allows one to map the ground state properties of (\ref{1}) to those of a classical 1D Ising model with long range interactions governed by the Fourier transform $G(\tau) = \int (d\omega/2\pi) J(\omega) e^{-|\omega| \tau} \sim 1/|\tau|^{1+s}$. For $s \leq 1$ (and $h_z = 0$) the interaction induces a phase transition to a ferromagnetic state $\langle \hat S_z \rangle \sim (\alpha - \alpha_c)^{\beta(s)}$ with increasing $\alpha > \alpha_c(h_x)$ and order parameter exponent $\beta(s)$.

For present purposes, detailed properties of this transition are unimportant, only that it exists (i.e., $s \leq 1$). As the first step we adiabatically turn on the environmental coupling, evolving it to final values $(h_x,\alpha)$ along a trajectory within the ferromagnetic phase, with result
\begin{equation}
|\Psi_U \rangle = \hat U(h_x,\alpha) |\Psi_0 \rangle,\ \
|\Psi_0 \rangle = |\psi_0 \rangle \otimes |\psi_B^\mathrm{gnd} \rangle,
\label{3}
\end{equation}
corresponding to adiabatic evolution implemented by a unitary operator $\hat U(h_x,\alpha)$ acting on the product of the initial spin state and boson ground state $|\psi_B^\mathrm{gnd} \rangle$. The entangled spin--boson state $|\Psi_U \rangle = C_+ |U,+ \rangle + C_- |U, -\rangle$ is a \emph{macroscopic} superposition (``cat state'') formed from two opposite ferromagnetic states $|U, \pm \rangle = \hat U(h_x,\alpha) |\pm \rangle \otimes |\psi_B^\mathrm{gnd} \rangle$. It is useful to think of first generating the state $\hat U(0,\alpha) |\Psi_0 \rangle$ at $h_x = 0$. Since $\hat S_z$ then commutes with $H_{SB}$ its eigenvalues $\sigma = \pm \frac{1}{2}$ are good quantum numbers, and the two ferromagnetic states, obtained by completing the square in (\ref{1}), correspond to opposite shifts $\tilde b_k = \hat b_k - \sigma \lambda_k/\omega_k$ in the harmonic oscillator origin and corresponding ground state energy reduction $\Delta E = -\sum_k \lambda_k^2/4\omega_k = -\int (d\omega/4\pi \omega) J(\omega)$. One may then turn on $h_x$, which induces tunneling between the spin states and corresponding quantum fluctuations in the oscillator origins. But so long as $\alpha > \alpha_c(h_x)$ the two states remain distinct \cite{foot:paraphase}.

\paragraph{Physical model example:} To gain intuition, consider a spin embedded in an elastic medium with real space form $H_{SB}' = -\hat S_z \int d{\bf r} \lambda({\bf r}) [\hat \psi({\bf r}) + \hat \psi^\dagger({\bf r})]$ with coupling $\lambda({\bf r})$ supported on a macroscopic, but physically small (e.g., volume $V = 1$ mm$^3$) ``detector pixel''. Using Fourier decomposition $\hat \psi({\bf r}) = V^{-1/2} \sum_{\bf k} \hat b_{\bf k} e^{-i{\bf k} \cdot {\bf r}}$, one obtains $\lambda_k \to \lambda({\bf k}) = V^{-1/2} \int d{\bf r} \lambda({\bf r}) e^{-i{\bf k} \cdot {\bf r}}$, and $H_B$ represents an elastic (phonon) energy with dispersion relation $\omega({\bf k}) \approx c_\mathrm{ph} |{\bf k}|$. The oscillator shift leads to a corresponding macroscopic real space density shift $\rho^0({\bf r}) = \langle \hat \psi^\dagger({\bf r}) \hat \psi({\bf r})\rangle = \sum_{\bf k} [\lambda({\bf k})/2\omega({\bf k})]^2 e^{i{\bf k} \cdot {\bf r}}$ within the pixel, and related to the Fourier transform of $J(\omega)/\omega^2$.

\paragraph{Open system measurement model:} The distinct states $|U, \pm\rangle$ exist only in the thermodynamic limit where the origin shift effectively generates an infinite mass displacement. This is the key ``classical'' many body response permitting robust measurement. At this stage, since we are still dealing with an ideal closed system, $|\Psi_U \rangle$ is a cat state fully encoding the original complex amplitudes $C_\pm$. However, even just preparing for an external measurement puts this state into contact with the external world. Motivated by stray EM interactions with the boson environment, we extend (\ref{1}) in the form $H_\mathrm{SBA} = H_\mathrm{SB} + H_A + H_{AB}^{(1)} + H_{AB}^{(2)}$ with additional terms
\begin{eqnarray}
H_A &=& \sum_l \Omega_l \hat a_l^\dagger \hat a_l,\ \
H_{AB}^{(1)} = \sum_{k,l,m} \lambda^{(1)}_{klm} (\hat a_k + \hat a_k^\dagger)
\hat b_l^\dagger \hat b_m
\nonumber \\
H_{AB}^{(2)} &=& \sum_{k,l,m,n} \lambda^{(2)}_{klmn} \hat a_k^\dagger \hat a_l
\hat b_m^\dagger \hat b_n,
\label{4}
\end{eqnarray}
with $H_A$ defining a new bosons $\hat a_l, \hat a_l^\dagger$. Here $H_{AB}^{(1,2)}$ are motivated by the usual photon--matter coupling terms: current coupling ${\bf J} \cdot {\bf A}$ and density coupling $\rho |{\bf A}|^2$, respectively \cite{Baym}. The new bath provides weak external illumination of $|\Psi_U \rangle$, making the components $|\pm, U \rangle$ ``visible'' to each other, and, as we shall see, providing a decoherence mechanism leading to wavefunction collapse.

Leaving a more complete analysis of (\ref{4}) for future work, we pursue here an effective mesoscale model approach, exhibiting the basic decoherence effects. We define a superspin operator $\hat {\bm \Sigma}$, with eigenstates $|\pm \rangle_\Sigma$ of $\hat \Sigma_z$ identified with the spin--boson states $|\pm, U \rangle$. We then summarize the higher level environmental interaction using Hamiltonian $H_\mathrm{\Sigma B} = H_\Sigma + \tilde H_B + H'_{\Sigma B}$ with terms
\begin{eqnarray}
H_\Sigma &=& -H_z \hat \Sigma_z - H_x \hat \Sigma_x,\ \
\tilde H_B = \sum_l \Omega_l \hat B_l^\dagger \hat B_l
\nonumber \\
H'_{\Sigma B} &=& -\hat \Sigma_z \sum_l
\Lambda_l (\hat B_l + \hat B_l^\dagger)
\label{5}
\end{eqnarray}
directly analogous to (\ref{1}). The full bath $\{\hat a_k\}$ is subsumed into a projected bath $\{\hat B_l\}$ that couples directly to $\hat \Sigma_z$, and an effective transverse field $H_x$ that induces tunneling between the states $|\pm \rangle_\Sigma$. The former emerges from $H_{AB}^{(1)}$, with $\Lambda_l \sim \langle \hat b_l^\dagger \hat b_l \rangle$ responding to the background distortion encoded in $|\pm, U \rangle$, and may be thought of as a long wavelength, low frequency filter of the $\{\hat a_k \}$. The latter emerges most directly from $H_{AB}^{(2)}$, governed at leading order by the matrix element $H_x \approx \langle +,U|H_{AB}^{(2)}|-,U \rangle$ \cite{foot:tunneling}. Estimates based on the optical model $H_{AB}^{(2)} = (e^2/2mc^2) \int d{\bf r} {\bf A}({\bf r})^2 \psi^\dagger({\bf r}) \psi({\bf r})$ lead to $H_x/\hbar \sim \frac{e^2}{\hbar c} \frac{k_B T}{mc^2} \frac{k_B T}{\hbar} N^0$ where $N^0 = \int d{\bf r} \rho^0({\bf r})$ is mean total boson particle displacement \cite{SuppMat}. Even using rather conservative estimates (e.g., fraction $10^{-9}$ electron cloud displacement within a 1 mm$^3$ ``pixel'' at $T = 1$ K), the tunneling rate $H_x/\hbar \sim 10^{10}$ 1/s is extremely large due to the macroscopic nature of the two states \cite{Zurek1986}.

\paragraph{Reduced density matrix and decoherence:} In order to study decoherence within the effective model (\ref{5}), we consider the evolution of the reduced density matrix
\begin{equation}
\hat \rho_\Sigma(t) = \mathrm{tr}^B\left[e^{-it H_{\Sigma B}}
\hat \rho_0 e^{it H_{\Sigma B}} \right]
\equiv \sum_{\sigma,\sigma' = \pm 1} A_{\sigma \sigma'}(t)
|\sigma \rangle \langle \sigma'|
\label{6}
\end{equation}
obtained by averaging over the boson degrees of freedom. The initial condition is $\hat \rho_0 = |\Psi_\Sigma \rangle \langle \Psi_\Sigma | \otimes \hat \rho^B_\mathrm{eq}$, with initial superspin state $|\Psi_\Sigma \rangle = C_+ |+ \rangle + C_- |- \rangle$ representing the macrostate $|\Psi_U \rangle$, and equilibrium distribution $\hat \rho^B_\mathrm{eq} = Z_A^{-1} e^{-\tilde H_B/T}$ at a temperature $T$ (with partition function $Z_A(T) = \mathrm{tr}^B[e^{-\tilde H_B/T}]$).

Although $H_x$ provides the critical coupling between the macrostates, its role is otherwise assumed small, providing a weak perturbation compared to $H'_{\Sigma B}$. The zeroth order exact solution is
\begin{equation}
A_{\sigma \sigma'}(t) = C_\sigma C_{\sigma'}^*
e^{-\delta_{\sigma, -\sigma'} {\cal D}(t)}
\label{7}
\end{equation}
with decoherence function defined by the $H_x = 0$ equilibrium average, ${\cal D}(t) = -\ln \langle e^{it H'_{\Sigma B}} \rangle_\mathrm{eq}$, yielding \cite{SuppMat}
\begin{equation}
{\cal D}(t) = 16 \int \frac{d\Omega}{\pi \Omega^2} {\cal J}(\Omega)
\left[n^T_B(\Omega) + \textstyle{\frac{1}{2}} \right] \sin^2(\Omega t/2)
\label{8}
\end{equation}
with Bose function $n^T_B(\Omega) = (e^{\Omega/T} -1)^{-1}$ and spectral function ${\cal J}(\Omega) = \pi \sum_l \Lambda_l^2 \delta(\Omega - \Omega_l) = 2\pi \alpha_\Sigma \Omega^S e^{-\Omega/\Omega_c}$. For the physically interesting range $0 < S < 2$, ${\cal D}(t) \approx -8 \alpha_\Sigma T \Gamma(S-2) \cos[\pi(S-2)/2] t^{2-S}$ at large $t \gg 1/\Omega_c, 1/T$, implying exponential decay on (thermal) time scale $t_\mathrm{coh} \sim (8\alpha_\Sigma T)^{-1/(2-S)}$ \cite{foot:T0decoh}. One therefore recovers the Born rule $\hat \rho_\Sigma(t \to \infty) = \sum_\sigma |C_\sigma|^2 |\sigma \rangle \langle \sigma |$. The linear in $H_x$ correction is of the form $\delta \hat \rho_\Sigma(t) = \chi_{\cal J}(t) H_x (|+ \rangle \langle -| + |- \rangle \langle +|)$ which acts to slightly mix the two states, quantified by transverse susceptibility \cite{SuppMat}
\begin{eqnarray}
\chi_{\cal J}(t) &=& 2\int_0^t ds e^{-{\cal D}(s)} \sin[\Phi(t,s)]
\nonumber \\
\Phi(t,s) &=& 16 \int \frac{d\Omega}{\pi \Omega^2} {\cal J}(\Omega)
\sin(\Omega t/2) \sin(\Omega s/t)
\nonumber \\
&&\times\ \sin[\Omega(t-s)/2],
\label{9}
\end{eqnarray}
while retaining the same probabilities $|C_\pm|^2$.

Although verifying the Born rule, the solution (\ref{7}) \emph{does not} directly prove a unique classical solution, with state $|\Psi_\Sigma(t) \rangle = e^{-it H_{\Sigma B}} |\Psi_\Sigma \rangle \otimes |\Psi^B \rangle_T$ converging to one of $|\pm \rangle$. Here $|\Psi^B_T \rangle = \otimes_l |n_l\rangle$ is a random realization of the thermal boson state, with independent Poisson distributed occupancies $n_l$ with mean $n^T_B(\Omega_l)$. In fact, for $H_x = 0$ the state $|\Psi_\Sigma(t) \rangle = \sum_\sigma C_\sigma e^{-it H_{\sigma B}} |\sigma \rangle \otimes |\Psi^B \rangle_T$ is still a macroscopic superposition, with $H_{\sigma B} = H_{\Sigma B}(\hat \Sigma_z \to \sigma)$. Even if $H_z \neq 0$, explicitly breaking the symmetry, so long as $H_x = 0$ there is an exact Hilbert space decomposition ${\cal H} = {\cal H}_+ \oplus {\cal H}_-$ into $\sigma = \pm 1$ sectors, with no communication between them. In particular, even as they interact with the same environmental degrees of freedom, the two sectors equilibrate separately. However, the resulting states do have hugely different phase factors, rapidly suppressing density matrix off-diagonal elements.

On the other hand, for $H_x \neq 0$ sector independence is broken, and one expects $\hat \rho_\Sigma(t)$ to reflect the expected classical outcome---the two macroscopic states are now ``visible'' to each other, and it is unlikely that the equilibration process can preserve both. Thus, for small $H_x$, a multi-Poisson choice for $|\Psi^B \rangle_T$ should lead to one of two late time states of the form $|\sigma \rangle \otimes |\Psi^B_\sigma(t) \rangle$ in which $|\Psi^B_\sigma(t) \rangle$ is a boson state that retains a fluctuating unitary dynamics encoding the original coefficients $C_\pm$, now ``lost'' to the environment. The density matrix (\ref{6}) then reflects a classical average over multiple experiments, each with a definite ``pointer state'' outcome. This result is directly analogous to standard phase transition results in open systems (classical or quantum): equilibration dynamics ultimately converges to one of the broken symmetry states (in general via some complex phase separation process---e.g., domain growth in Ising systems or vortex tangle decay in superfluid systems), depending on precise initial condition. Verifying this here would be an extremely interesting topic for future investigation. The model (\ref{5}) is sufficiently simple that treating the random initial condition problem should be tractable.

\paragraph{Multiple entangled spins:} We end with an application to physically separated entangled spins. The previous model results are adapted to provide explicit predictions for measurements on all or a subset spins, fully verifying conventional Born rule assumptions. The model explicitly exhibits subtleties related to information loss to the environment. We consider here only two spins. Generalization to a greater number is straightforward.

Consider an entangled initial state $|\psi_0^{(1,2)} \rangle = \sum_{\sigma,\sigma'} C_{\sigma \sigma'} |\sigma \rangle_1 \otimes |\sigma' \rangle_2$. The measurement is performed independently on each spin. Individual spin measurement outcomes are governed by the reduced density matrices
\begin{equation}
\hat \rho^S_m = \mathrm{tr}_{\bar m}
\left[|\psi_0^{(1,2)} \rangle \langle \psi_0^{(1,2)}| \right]
= \sum_{\sigma,\sigma'} A^{(m)}_{\sigma \sigma'}
|\sigma \rangle_m {}_m\langle \sigma'|
\label{10}
\end{equation}
in which $\bar m = 3-m$, $m=1,2$, is the complementary spin label, yielding $A^{(1)}_{\sigma \sigma'} = [{\bf C} {\bf C}^\dagger]_{\sigma \sigma'} = \sum_{\sigma_2} C_{\sigma \sigma_2} C^*_{\sigma' \sigma_2}$, $A^{(2)}_{\sigma \sigma'} = [{\bf C}^\dagger {\bf C}]_{\sigma' \sigma} = \sum_{\sigma_1} C_{\sigma_1 \sigma} C^*_{\sigma_1 \sigma'}$. Conventional treatment of the measurement process is to again assume decoherence of the off-diagonal terms, leaving the diagonal terms $p_\sigma^{(m)} = A_{\sigma\sigma}^{(m)}$ to define the final Born rule probabilities for independent measurements of each spin.

Consistently, we now show that the measurement protocol applied separately to each spin generates one of the four the pointer states $|U,\sigma \rangle_1 \otimes |U, \sigma'\rangle_2$ with classical probability $p_{\sigma \sigma'} = p_\sigma^{(1)} p_{\sigma'}^{(2)}$. Critically, these probabilities reflect properties only of the original entanglement operation, consistent with no further transfer of information between the subsystems during the measurement.

The measurement result on a given system is summarized by a unitary operation $\hat U_\mathrm{meas}(h_x,\alpha) = \hat U_\mathrm{SBA} \hat U(h_x,\alpha)$, with $\hat U(h_x,\alpha)$ the adiabatic evolution producing $|\pm, U \rangle$ derived from (\ref{1}), and $\hat U_\mathrm{SBA} = e^{-it_\mathrm{eq} H_\mathrm{SBA}}$ derived from (\ref{4}) for some sufficiently large equilibration/decoherence time $t_\mathrm{eq}$. First, the measurement performed only on subsystem 2 yields density matrix
\begin{equation}
\hat \rho^{(1,2)} = \hat U^{(2)}_\mathrm{meas}(h_{x,2},\alpha_2)
|\Psi_{SBA}^{(1,2)} \rangle \langle \Psi_{SBA}^{(1,2)}|
\hat U^{(2)}_\mathrm{meas}(h_{x,2},\alpha_2)^\dagger,
\label{11}
\end{equation}
in which $|\Psi_{SBA} \rangle = |\psi_0^{(1,2)} \rangle \otimes |\Psi_B^\mathrm{gnd} \rangle_1 \otimes |\Psi^A_T \rangle_1 \otimes  |\Psi_B^\mathrm{gnd} \rangle_2 \otimes |\Psi^A_T \rangle_2$ is the initial dual system spin--environment product state, with entanglement only in the spin. Using the cyclic property of the trace, one obtains
\begin{eqnarray}
\hat \rho^{(1)} &\equiv& \mathrm{tr}_2[\hat \rho^{(1,2)}]
= \mathrm{tr}_2[|\Psi_{SBA} \rangle \langle \Psi_{SBA}|]
\nonumber \\
&=& \hat \rho_1^S
\otimes (|\psi_{B,1}^\mathrm{gnd} \rangle \langle \psi_{B,1}^\mathrm{gnd}|)
\otimes (|\Psi_{A,1}^T \rangle \langle \Psi_{A,1}^T|),\ \ \ \ \ \
\label{12}
\end{eqnarray}
consisting of a direct product of the original spin and subsystem 1 boson density matrices. It is critical that the trace include the higher level thermal subsystem 2 boson degrees of freedom: these encode the information ``lost'' during equilibration. If one were to instead substitute one of the pointer states $|\pm, U \rangle_2$, the state choice impacts the result for $\hat \rho_1^S$, and one mistakenly concludes that entanglement with subsystem 2 has impacted the state of ${\bf S}_1$ in the absence of physical interaction.

If, following equilibration, one traces out both sets of thermal boson degrees of freedom, one arrives instead at the spin--boson state density matrix
\begin{equation}
\hat \rho^{(1,2)}_\mathrm{SB}
= \hat \rho^S_1 \otimes (|\psi_{B,1}^\mathrm{gnd} \rangle
\langle \psi_{B,1}^\mathrm{gnd}|)
\otimes (|\sigma_2, U \rangle_2 {}_2\langle\sigma_2, U|),
\label{13}
\end{equation}
for randomly chosen value $\sigma_2$ (with probability $p_{\sigma_2}^{(2)}$ inferred as before from the equilibrium thermal average). Following the thermal average, the only entanglement information remaining resides in the coefficients $A_{\sigma \sigma'}^{(1)}$ defining $\hat \rho^S_1$. The subsystem 2 state depends on the adiabatic endpoints $h_{x,2},\alpha_2$. For a measurement performed on both subsystems, in arbitrary order, one recovers the spin--boson product state $\hat \rho^{(1,2)}_\mathrm{SB} = (|\sigma_1,U \rangle_1 {}_1\langle \sigma_1, U|) \otimes (|\sigma_2, U \rangle_2  {}_2\langle \sigma_2, U|)$ with desired probability $p_{\sigma_1 \sigma_2}$.

The measurements so far encompass only single spin averages, hence fail to explore more interesting aspects of entanglement. For example, the correlation $G_{zz} = \langle \psi_0| \hat S^{(1)}_z \hat S^{(2)}_z |\psi_0 \rangle = \frac{1}{4} \sum_{\sigma,\sigma'} \sigma \sigma' |C_{\sigma \sigma'}|^2$ generally differs from the product of the averages $M_{zz} = \langle \hat S_z^{(1)}(\alpha_1) \rangle_1 \langle \hat S_z^{(2)}(\alpha_2) \rangle_2 = [\sum_\sigma \sigma p^{(1)}_\sigma] [\sum_{\sigma'} \sigma' p^{(2)}_{\sigma'}]$. For the antisymmetric state $C_{\sigma \sigma'} = 2^{-1/2} \sigma \delta_{\sigma,-\sigma'}$ one obtains $G_{zz} = -\frac{1}{4}$, $M_{zz} = 0$, indicating mean zero, but perfectly anticorrelated spins. Moreover, the probabilities $p_\sigma^{(m)} = \frac{1}{2}$ do not distinguish, e.g., the antisymmetric state from the symmetric state $C_{\sigma \sigma'} = 2^{-1/2} \delta_{\sigma,-\sigma'}$.

To measure multispin correlations, one must couple both spins to the same boson degrees of freedom, e.g., via an environmental coupling of the form
\begin{equation}
H'_\mathrm{SB} = -\hat S_z^{(1)} \hat S_z^{(2)}
\sum_k \lambda_k (\hat b_k + \hat b_k^\dagger)
\label{14}
\end{equation}
in (\ref{1}). How to actually design such a measurement is an interesting problem. Consistent with quantum rules, and exhibited explicitly via the protocol developed here, only one measurement sequence is permitted. Additional sequences require a newly prepared system.

\appendix

\section{Supplementary Material}

This Appendix contains some detailed derivations of results summarized in the main text. Section \ref{app:dmevol} provides details of the leading order density matrix evolution calculations leading to the leading order decoherence result summarized in equation (8) in the main text. Section \ref{app:dmHx} generalizes these results to include corrections for finite transverse field $H_x$ (which induces tunneling between the spin-Boson ground states), leading to the linear correction summarized in equation (9) of the main text. Section \ref{app:parmest} provides details of the optical model estimate for the tunneling rate described in the paragraph below equation (5) of the main text, derived from the microscopic model, equation (4) in the main text. The results lead to physical estimates for the parameter $H_x$ in the effective model Hamiltonian $H_{\Sigma B}$.

\subsection{Density matrix evolution and decoherence time}
\label{app:dmevol}

We seek to evaluate a time-dependent density matrix of the form
\begin{eqnarray}
\hat \rho_{w w'}(t) &=& e^{-i t H_w} e^{-\beta H_0} e^{i t H_{w'}}
\nonumber \\
&=& \hat \rho_{w w}(t) e^{-i t H_w} e^{i t H_{w'}}
\label{A1}
\end{eqnarray}
in which $\beta = 1/T$ is the inverse temperature, and
\begin{equation}
H_w = \Omega (\hat B^\dagger - w^*) (\hat B - w)
\label{A2}
\end{equation}
where $w$ is an arbitrary complex number \cite{foot:rho}. Defining the operator,
\begin{equation}
\hat P_w = i (w^* \hat B - w \hat B^\dagger),
\label{A3}
\end{equation}
one may derive the translation identity
\begin{equation}
e^{i\hat P_w} \hat B e^{-i \hat P_w} = \hat B - w,
\label{A4}
\end{equation}
from which it follows that
\begin{equation}
H_w = e^{i\hat P_w} H_0 e^{-i\hat P_w}
\label{A5}
\end{equation}
and hence
\begin{eqnarray}
\hat \rho_{w w}(t) &=& e^{i\hat P_w} e^{-i t H_0} e^{-i \hat P_w}
e^{-\beta H_0} e^{i \hat P_w} e^{i t H_0} e^{-i \hat P_w}
\nonumber \\
e^{-i t H_w} e^{i t H_{w'}} &=& e^{i\hat P_w} e^{-i t H_0} e^{-i\hat P_w}
e^{i\hat P_{w'}} e^{i t H_0} e^{-i\hat P_{w'}}.
\nonumber \\
\label{A6}
\end{eqnarray}
The standard harmonic oscillator solution \cite{Merzbacher} implies that
\begin{equation}
e^{-it H_0} \hat B e^{it H_0} = \hat B e^{i\Omega t},
\label{A7}
\end{equation}
and one therefore obtains
\begin{eqnarray}
e^{-i t H_0} e^{-i \hat P_w}
e^{-\beta H_0} e^{i \hat P_w} e^{i t H_0}
&=& e^{-i\hat P_w(t)}
e^{-\beta H_0} e^{i \hat P_w(t)}
\nonumber \\
e^{-i t H_0} e^{-i\hat P_w} e^{i\hat P_{w'}} e^{i t H_0}
&=& e^{-i \hat P_w(t)} e^{i\hat P_{w'}(t)}
\label{A8}
\end{eqnarray}
in which
\begin{equation}
\hat P_w(t) = e^{-i t H_0} \hat P_w e^{i t H_0}
= i(w^* \hat B e^{i\Omega t} - w \hat B^\dagger e^{-i\Omega t}).
\label{A9}
\end{equation}
From the identity \cite{Merzbacher}
\begin{equation}
e^{A+B} = e^A e^B e^{-\frac{1}{2}[A,B]},
\label{A10}
\end{equation}
valid whenever $[A,B]$ is a c-number (or more generally, whenever $[A,B]$ commutes with both $A$ and $B$), it follows that
\begin{eqnarray}
e^{i\hat P_w} e^{i\hat P_{w'}} &=& e^{i\hat P_{w+w'}}
e^{-\frac{1}{2}[\hat P_w,\hat P_{w'}]}
\nonumber \\
{[}\hat P_w,\hat P_{w'}] &=& w^* w' - w w^{\prime *}.
\label{A11}
\end{eqnarray}
Using (\ref{A11}) one may simplify (\ref{A6}) to the form
\begin{eqnarray}
\hat \rho_{ww}(t) &=& e^{i[\hat P_w - \hat P_w(t)]}
e^{-\beta H_0} e^{-i [\hat P_w - \hat P_w(t)]}
\nonumber \\
&=& e^{-\beta H_{w(t)}}
\nonumber \\
e^{-i t H_w} e^{i t H_{w'}} &=& e^{i \hat P_{w(t) - w'(t)}}
e^{\frac{i}{2} C_{ww'}(t)}
\label{A12}
\end{eqnarray}
in which we define
\begin{eqnarray}
w(t) &=& w(1 - e^{-i\Omega t})
\nonumber \\
C_{ww'}(t) &=& \frac{1}{i}\left\{[\hat P_w, \hat P_w(t)]
- [\hat P_{w'}, \hat P_{w'}(t)] \right.
\nonumber \\
&&+\ \left. [\hat P_{w(t)}, \hat P_{w'(t)}] \right\}
\nonumber \\
&=& 2 (|w'|^2 - |w|^2) \sin(\Omega t)
\nonumber \\
&&-\ 4i(w^*w' - w w^{\prime *}) \sin^2(\Omega t/2).
\label{A13}
\end{eqnarray}
The definition ensures that $C_{ww'}(t)$ is real. For the special case $w' = -w$ one obtains $\hat P_{w'} = -\hat P_w$, and hence
\begin{equation}
C_{w,-w}(t) \equiv 0.
\label{A14}
\end{equation}

\subsubsection{Equilibrium averages}
\label{app:eqave}

The cyclic property of the trace yields the trivial result
\begin{eqnarray}
Z_0 &\equiv& \mathrm{tr}[\hat \rho_{ww}(t)]
= \mathrm{tr}[e^{-\beta H_0}]
\nonumber \\
&=& \frac{1}{1-e^{-\beta \Omega}} \equiv e^{-\beta F_0}
\nonumber \\
F_0 &=& \frac{1}{\beta} \ln(1 - e^{-\beta \Omega}).
\label{A15}
\end{eqnarray}
For the general case one obtains by applying the cyclic property to the first line of (\ref{A1}), and then using the second line of (\ref{A12}),
\begin{eqnarray}
\mathrm{tr}[\hat \rho_{ww'}(t)] &=&
\mathrm{tr}\left[e^{-\beta H_0} e^{it H_{w'}} e^{-itH_w} \right]
\nonumber \\
&=& e^{\frac{1}{2} C_{w'w}(-t)}
\mathrm{tr}\left[e^{-\beta H_0} e^{i \hat P_{w'(-t) - w(-t)}} \right]
\nonumber \\
&=&  e^{\frac{1}{2} C_{w'w}(-t)} \sum_{n=0}^\infty e^{-n \beta \Omega}
\langle n| e^{i \hat P_{w'(-t) - w(-t)}} |n \rangle.
\nonumber \\
\label{A16}
\end{eqnarray}
To evaluate the matrix element, we use (\ref{A10}) to express
\begin{equation}
e^{i \hat P_w} = e^{wB^\dagger} e^{-w^*B} e^{-\frac{1}{2} |w|^2},
\label{A17}
\end{equation}
which then yields
\begin{eqnarray}
\langle n| e^{i \hat P_w} |n \rangle
&=& e^{-\frac{1}{2} |w|^2} \langle n |e^{wB^\dagger} e^{-w^*B} |n \rangle
\nonumber \\
&=& e^{-\frac{1}{2} |w|^2} \sum_{m=0}^n \frac{(-1)^m |w|^{2m}}{(m!)^2}
\langle n |({\hat B}^\dagger)^m {\hat B}^m |n \rangle
\nonumber \\
&=& e^{-\frac{1}{2} |w|^2} \sum_{m=0}^n
\frac{(-1)^m n!}{(m!)^2 (n-m)!} |w|^{2m}.
\label{A18}
\end{eqnarray}
Making use of the identity
\begin{equation}
\sum_{l=0}^\infty \frac{(l+m)!}{l!} x^l
= \frac{d^m}{dx^m} \frac{1}{1-x}
= \frac{m!}{(1-x)^{m+1}}
\label{A19}
\end{equation}
one then obtains
\begin{eqnarray}
\langle e^{i \hat P_w} \rangle
&\equiv& \frac{1}{Z_0} \sum_{n=0}^\infty e^{-n \beta \Omega}
\langle n| e^{i \hat P_w} |n \rangle
\nonumber \\
&=& e^{-\frac{1}{2} |w|^2} \frac{1}{Z_0}
\sum_{m=0}^\infty \frac{(-1)^m |w|^{2m}}{(m!)^2}
\nonumber \\
&&\times\ \sum_{n=m}^\infty \frac{n!}{(n-m)!} e^{-n \beta \Omega}
\nonumber \\
&=& e^{-\frac{1}{2} |w|^2} \sum_{m=0}^\infty
\frac{(-1)^m |w|^{2m}}{m!}
\frac{e^{-m\beta \Omega}}{(1-e^{-\beta \Omega})^m}
\nonumber \\
&=& e^{-\left[n_B(\Omega) + \frac{1}{2} \right] |w|^2}
\label{A20}
\end{eqnarray}
where
\begin{equation}
n_B(\Omega) = \frac{1}{e^{\beta \Omega} - 1}
\label{A21}
\end{equation}
is the equilibrium Bose occupation number. Finally, substituting this result back into (\ref{A16}), one obtains
\begin{eqnarray}
\frac{1}{Z_0} \mathrm{tr}[\hat \rho_{ww'}(t)]
&=& \left\langle e^{iH_w t} e^{-i H_{w'} t} \right\rangle
\nonumber \\
&=& e^{\frac{i}{2} C_{w'w}(-t)} e^{-D_{ww'}(t)},
\label{A22}
\end{eqnarray}
in which
\begin{eqnarray}
D_{ww'}(t) &=& \left[n_B(\Omega) + \frac{1}{2} \right] |w'(-t) - w(-t)|^2
\nonumber \\
&=& 4 \left[n_B(\Omega) + \frac{1}{2} \right] |w'-w|^2 \sin^2(\Omega t/2)
\nonumber \\
C_{w'w}(-t) &=& 2 (|w'|^2 - |w|^2) \sin(\Omega t)
\nonumber \\
&&+\ 4i(w^*w' - w w^{\prime *}) \sin^2(\Omega t/2)
\label{A23}
\end{eqnarray}
are both real quantities. For the special case $w' = -w$ this simplifies to
\begin{equation}
\frac{1}{Z_0} \mathrm{tr}[\hat \rho_{w,-w}(t)]
= e^{-16 \left[n_B(\Omega) + \frac{1}{2} \right] |w|^2 \sin^2(\Omega t/2)}.
\label{A24}
\end{equation}

The multi-Boson solution is simply the product over $l$ of the result (\ref{A24}), substituting $\Omega \to \Omega_l$ and $w \to \Lambda_l/\Omega_l$. This form is obtained by completing the square in the effective model $H_{\Sigma B} = \tilde H_B + H'_{\Sigma B}$, equation (5) in the main text (with $H_\Sigma = 0$). The result is precisely equation (8) in the main text.

\subsubsection{Decoherence time}
\label{sec:tdecoherence}

Explicit evaluation of ${\cal D}(t)$ makes use of the function
\begin{eqnarray}
F_p(\tau) &=& 4 \int_0^\infty du u^{p-1} e^{-u} \sin^2(\tau u/2)
\nonumber \\
&=& \Gamma(p) \left[2 - \frac{1}{(1-i\tau)^p} - \frac{1}{(1+i\tau)^p} \right],
\ \ \ \ \ \
\label{A25}
\end{eqnarray}
valid when the integral converges at the origin, $p > -2$. The apparent singularities at $p = 0,-1$ actually generate logarithmic behavior:
\begin{eqnarray}
F_0(\tau) &=& \ln(1+\tau^2)
\nonumber \\
F_{-1}(\tau) &=& 2\tau \mathrm{arctan}(\tau) - \ln(1+\tau^2).
\label{A26}
\end{eqnarray}
The large $|\tau| \gg 1$ asymptotic limit yields
\begin{equation}
F_p(\tau) \approx \left\{\begin{array}{ll}
-2 \Gamma(p) \cos\left(\frac{\pi p}{2} \right) |\tau|^{|p|},
& -2 < p < 0,\ p \neq -1 \\
\pi |\tau| - 2\ln|\tau|, & p = -1 \\
2 \ln|\tau|, & p = 0 \\
2 \Gamma(p), & p > 0.
\end{array} \right.
\label{A27}
\end{equation}

Inserting the model form
\begin{equation}
{\cal J}(\Omega) = \alpha_\Sigma \Omega^S e^{-\Omega/\Omega_c},
\label{A28}
\end{equation}
into the decoherence function, equation (8) in the main text, one may separate ${\cal D}(t)$ into zero and finite temperature contributions,
\begin{equation}
{\cal D}(t) = {\cal D}_0(t) + {\cal D}_B(t),
\label{A29}
\end{equation}
in which the Bose function contribution is
\begin{eqnarray}
{\cal D}_B(t) &=& 16 \alpha_\Sigma \int_0^\infty d\Omega \Omega^{S-2}
e^{-\Omega/\Omega_c} n_B(\Omega) \sin^2(\Omega t/2)
\nonumber \\
&\approx& \frac{16 \alpha_\Sigma}{\beta} \int_0^\infty d\Omega \Omega^{S-3}
e^{-\Omega/\Omega_c} \sin^2(\Omega t/2),\ \ t \gg \beta
\nonumber \\
&=& \frac{4 \alpha_\Sigma \Omega_c^{S-2}}{\beta} F_{S-2}(\Omega_c t),
\label{A30}
\end{eqnarray}
valid for $S > 0$. On the other hand, for $t \ll \beta$ (low temperature regime) one may neglect the $n_B(\Omega)$ term and obtain the exact form
\begin{eqnarray}
{\cal D}_0(t) &=& 8 \alpha_\Sigma \int_0^\infty
d\Omega \Omega^{S-2}e^{-\Omega/\Omega_C} \sin^2(\Omega t/2)
\nonumber \\
&=& 2\alpha_\Sigma \Omega_c^{S-1} F_{S-1}(\Omega_c t),
\label{A31}
\end{eqnarray}
valid for $S > -1$. Using (\ref{A25}) one may summarize the large time behavior in the form
\begin{widetext}
\begin{eqnarray}
{\cal D}_B(t) &\approx& \frac{8 \alpha_\Sigma}{\beta}
\left\{\begin{array}{ll}
-\Gamma(S-2) \cos[\pi(S-2)/2] t^{2-S}, & 0 < S < 2 \\
\ln(\Omega_c t), & S = 2 \\
\Gamma(S-2) \Omega_c^{S-2}  & S > 2
\end{array} \right.
\nonumber \\
{\cal D}_0(t) &\approx& 4 \alpha_\Sigma
\left\{\begin{array}{ll}
-\Gamma(S-1) \cos[\pi(S-1)/2] t^{1-S}, & -1 < S < 1 \\
\ln(\Omega_c t), & S = 1 \\
\Gamma(S-1) \Omega_c^{S-1} & S > 1
\end{array} \right.
\label{A32}
\end{eqnarray}
\end{widetext}
These are the results quoted below equation (8) and in footnote [10] in the main text. The asymptotic form for ${\cal D}_B(t)$ requires both $\Omega_c t \gg 1$ and $t/\beta \gg 1$, while that for ${\cal D}_0(t)$ requires only $\Omega_c t \gg 1$ and will dominate ${\cal D}_B(t)$ under the additional condition $t/\beta \ll 1$.

From (\ref{A32}) it follows that at finite temperature one obtains exponential decay of the off-diagonal terms (full decoherence) for $0 < S < 2$, with time scale
\begin{equation}
t_\mathrm{coh} \sim \left(\frac{\beta}{8\alpha_\Sigma} \right)^{1/(2-S)}.
\label{A33}
\end{equation}
Power law decay $\sim (\Omega_c t)^{-2\beta \alpha_\Sigma}$ is obtained for $S = 2$, and finite decoherence for $S > 2$. At zero temperature, exponential decay is obtained for $-1 < S < 1$, with time scale
\begin{equation}
t_{0,\mathrm{coh}} \sim (4\alpha_\Sigma)^{-1/(1-S)}.
\label{A34}
\end{equation}
Power law decay $\sim (\Omega_c t)^{-\alpha_\Sigma}$ is obtained for $S = 1$, and finite decoherence for $S > 1$. Since the states $|U,\pm \rangle$ represent macroscopic configurations of the original spin-boson system, one expects $\alpha_\Sigma$ to be macroscopic as well. A proper characterization would require a quantitative derivation of the relation between the microscopic model, equation (4) in the main text, and the effective reduced model, equation (5) in the main text, however one might expect $\alpha_\Sigma \sim N^0$ where
\begin{equation}
N^0 = \int d{\bf r} \rho^0({\bf r})
\label{A35}
\end{equation}
is a measure of the physical ``distance'' between the states $|\pm, U \rangle$, in this case the total number of displaced particles. Here the spin--boson spatial density response profile $\rho^0$ was defined in the \emph{Physical model example} section of the main text. Given its macroscopic character, one expects extremely rapid decoherence, likely shorter than any experimentally resolvable time \cite{Zurek1986a}.

\subsection{Density matrix perturbation theory for finite $H_x$}
\label{app:dmHx}

We consider next the leading corrections to the reduced density matrix for finite transverse field $H_x$, derived from formal perturbation theory based on the interaction picture representation. The main result is the off-diagonal density matrix correction, equation (9) in the main text.

\subsubsection{Interaction picture evolution of the transverse spin}
\label{app:ipsx}

We consider the interaction picture evolution of the operator $\hat \Sigma_x$ defined by \cite{FW1971}
\begin{equation}
\hat \Sigma_{x,I}(t) = e^{-it H_{w\Sigma}} \hat \Sigma_x e^{it H_{w\Sigma}}
\label{P1}
\end{equation}
in which [see (\ref{A2}) and (\ref{A3})]
\begin{eqnarray}
H_{w\Sigma} &=& \Omega \left(\hat B^\dagger - w^* \hat \Sigma_z \right)
\left(\hat B - w \hat \Sigma_z \right)
\nonumber \\
&=& H_0 - \hat \Sigma_z \hat P_{i\Omega w} + \Omega |w|^2
\label{P2}
\end{eqnarray}
for arbitrary complex number $w$, and we have recalled the normalizations $\hat \Sigma_\alpha^2 = 1$. The result is essentially a rotation about the $z$-axis that depends on the boson state.

Since $\hat \Sigma_x$ commutes with the boson operators, one may generalize the translation property (\ref{A4}) in the form
\begin{equation}
e^{i\hat \Sigma_z \hat P_w} \hat B e^{-i\hat \Sigma_z \hat P_w}
= \hat B - w \hat \Sigma_z,
\label{P3}
\end{equation}
and it follows as well that
\begin{equation}
H_{w\Sigma} = e^{i\hat \Sigma_z \hat P_w} H_0 e^{-i\hat \Sigma_z \hat P_w}.
\label{P4}
\end{equation}
One may therefore express
\begin{equation}
\hat \Sigma_{x,I}(t) = e^{i\hat \Sigma_z \hat P_w} e^{-it H_0}
e^{-i\hat \Sigma_z \hat P_w} \hat \Sigma_x e^{i\hat \Sigma_z \hat P_w}
e^{it H_0} e^{-i\hat \Sigma_z \hat P_w}.
\label{P5}
\end{equation}
Again, since the spin and boson operators commute, one may formally write
\begin{equation}
e^{i\hat \Sigma_z \hat P_w} = \cos(\hat P_w) + i \sin(\hat P_w) \hat \Sigma_z.
\label{P6}
\end{equation}
Using the identity \cite{Merzbacher}
\begin{equation}
[\hat \Sigma_\alpha, \hat \Sigma_\beta]
= 2i \epsilon_{\alpha \beta \gamma} \hat \Sigma_\gamma,
\label{P7}
\end{equation}
in which $\epsilon_{\alpha \beta \gamma}$ is the totally antisymmetric symbol with $\epsilon_{xyz} = 1$, one therefore obtains
\begin{eqnarray}
e^{-i\hat \Sigma_z \hat P_w} \hat \Sigma_x e^{i\hat \Sigma_z \hat P_w}
&=& \cos(2\hat P_w) \hat \Sigma_x + \sin(2\hat P_w) \hat \Sigma_y
\nonumber \\
&=& e^{2i \hat P_w} \hat \Sigma_- + e^{-2i \hat P_w} \hat \Sigma_+
\nonumber \\
e^{-i\hat \Sigma_z \hat P_w} \hat \Sigma_y e^{i\hat \Sigma_z \hat P_w}
&=& \cos(2\hat P_w) \hat \Sigma_y - \sin(2\hat P_w) \hat \Sigma_x
\nonumber \\
&=& e^{2i \hat P_w} i \hat \Sigma_- - e^{-2i \hat P_w} \hat i \Sigma_+
\nonumber \\
e^{-i\hat \Sigma_z \hat P_w} \hat \Sigma_\pm e^{i\hat \Sigma_z \hat P_w}
&=& e^{\mp 2i \hat P_w} \hat \Sigma_\pm
\label{P8}
\end{eqnarray}
with spin raising and lower operators $\hat \Sigma_\pm = \frac{1}{2}(\hat \Sigma_x \pm i \hat \Sigma_y)$. Inserting (\ref{P8}) into (\ref{P5}) one obtains
\begin{eqnarray}
\hat \Sigma_{x,I}(t) &=& e^{i\hat \Sigma_z \hat P_w}
[e^{2i\hat P_{w_t}} \hat \Sigma_-
+ e^{-2i\hat P_{w_t}} \hat \Sigma_+]
e^{-i\hat \Sigma_z \hat P_w}
\nonumber \\
&=& e^{i\hat \Sigma_z \hat P_w} e^{2i\hat P_{w_t}}
e^{-i\hat \Sigma_z \hat P_w}
(e^{i\hat \Sigma_z \hat P_w} \hat \Sigma_- e^{-i\hat \Sigma_z \hat P_w})
\nonumber \\
&&+\ e^{i\hat \Sigma_z \hat P_w} e^{-2i\hat P_{w_t}}
e^{-i\hat \Sigma_z \hat P_w}
(e^{i\hat \Sigma_z \hat P_w} \hat \Sigma_+ e^{-i\hat \Sigma_z \hat P_w})
\nonumber \\
&=& e^{i\hat \Sigma_z \hat P_w} e^{2i\hat P_{w_t}}
e^{-i\hat \Sigma_z \hat P_w} e^{-2i\hat P_w} \hat \Sigma_-
\nonumber \\
&&+\ e^{i\hat \Sigma_z \hat P_w} e^{-2i\hat P_{w_t}}
e^{-i\hat \Sigma_z \hat P_w} e^{2i\hat P_w} \hat \Sigma_+
\label{P9}
\end{eqnarray}
in which (\ref{A7}) has been used to evaluate the $e^{iH_0 t}$ evolution, and $w_t = w e^{-i\Omega t}$. In the last line (\ref{P8}) has been used again to evaluate the quantities in parentheses.  Finally, repeated use of the identities (\ref{A10}) and (\ref{A11}) may be used to combine the remaining $\hat P$ operators into single exponent
\begin{eqnarray}
\hat \Sigma_{x,I}(t) &=& e^{2i (\hat P_{w_t} - \hat P_w)}
e^{2(1 + \hat \Sigma_z)[\hat P_w,\hat P_{w_t}]} \hat \Sigma_-
\nonumber \\
&&+\ e^{-2i (\hat P_{w_t} - \hat P_w)}
e^{2(1 - \hat \Sigma_z)[\hat P_w,\hat P_{w_t}]} \hat \Sigma_+
\nonumber \\
&=& e^{-2i \hat P_{w(t)}} \hat \Sigma_-
+ e^{2i \hat P_{w(t)}} \hat \Sigma_+,
\label{P10}
\end{eqnarray}
where $w(t) = w - w_t = w(1 - e^{-i\Omega t})$ is the same as defined in (\ref{A13}), and cancellation of the commutator term follows from the property $f(\hat \Sigma_z) \hat \Sigma_\pm = f(\pm 1) \hat \Sigma_\pm$.

We can use the result (\ref{A20}) to compute equilibrium averages. The simplest example is
\begin{eqnarray}
\langle \hat \Sigma_{x,I}(t) \rangle
&=& \langle e^{-2i \hat P_{w(t)}} \rangle \hat \Sigma_-
+ \langle e^{2i \hat P_{w(t)}} \rangle \hat \Sigma_+
\nonumber \\
&=& e^{-4\left[n_B(\Omega) + \frac{1}{2} \right]|w(t)|^2}
(\hat \Sigma_- + \hat \Sigma_+)
\nonumber \\
&=& e^{-16\left[n_B(\Omega) + \frac{1}{2} \right] |w|^2 \sin^2(\Omega t/2)} \hat \Sigma_x
\label{P11}
\end{eqnarray}
which may be compared to (\ref{A24}). The second order dynamical correlation function is given by
\begin{eqnarray}
\langle \hat \Sigma_{x,I}(t) \hat \Sigma_{x,I}(t') \rangle
&=& \langle e^{-2i \hat P_{w(t)}} e^{2i \hat P_{w(t')}} \rangle
\hat \Sigma_- \hat \Sigma_+
\nonumber \\
&&+\ \langle e^{2i \hat P_{w(t)}} e^{-2i \hat P_{w(t')}} \rangle
\hat \Sigma_+ \hat \Sigma_-
\nonumber \\
&=& e^{-16 \left[n_B(\Omega) + \frac{1}{2} \right] |w|^2 \sin^2[\Omega(t-t')/2]}
\nonumber \\
&\times& e^{i 16 |w|^2 \sin[\Omega(t-t')/2] \sin(\Omega t/2) \sin(\Omega t'/2)}
\nonumber \\
\label{P12}
\end{eqnarray}
in which we note that $\hat \Sigma_\pm^2 = 0$ and $\hat \Sigma_- \hat \Sigma_+ + \hat \Sigma_+ \hat \Sigma_- = \openone$. Note that (\ref{P12}) is not a function of $t-t'$ alone because the time evolution (\ref{P1}) is with respect to a different Hamiltonian than that entering the equilibrium operator $e^{-\beta H_0}$.

\subsubsection{Many boson generalization}
\label{app:manybosongen}

Finally, we consider the generalization to multiple boson degrees of freedom in which the interaction picture evolution
\begin{eqnarray}
\hat \Sigma_{x,I}(t) = e^{-it H_{\Sigma B}^0} \hat \Sigma_x e^{it H_{\Sigma B}^0}
\label{P13}
\end{eqnarray}
is governed by the operator
\begin{eqnarray}
H_{\Sigma B}^0 &=& \tilde H_B + H'_{\Sigma B}
\nonumber \\
&=& \sum_l (H_{w_l \Sigma} - \Omega_l |w_l|^2)
\label{P14}
\end{eqnarray}
with definitions
\begin{eqnarray}
H_{w_l \Sigma} &=& \Omega_l (\hat B_l^\dagger - w_l \hat \Sigma_z)
(\hat B_l - w_l \hat \Sigma_z)
\nonumber \\
w_l &=& \frac{\Lambda_l}{\Omega_l},
\label{P15}
\end{eqnarray}
obtained from the effective Hamiltonian, equation (5) in the main text, with vanishing external field, $H_\Sigma = 0$. Since the boson operators for different $l$ all commute with each other, the result (\ref{P10}) immediately yields
\begin{eqnarray}
\hat \Sigma_{x,I}(t) &=& e^{-2i \hat P(t)} \hat \Sigma_-
+ e^{2i \hat P(t)} \hat \Sigma_+
\nonumber \\
\hat P(t) &=& \sum_l \hat P_{w_l(t)}
\label{P16} \\
&=& \sum_l \Lambda_l \frac{\sin(\Omega_l t/2)}{\Omega_l/2}
\left(e^{i\Omega_l t/2} \hat B_l
+ e^{-i\Omega_l t/2} \hat B_l^\dagger \right).
\nonumber
\end{eqnarray}

Using (\ref{A35}), the many boson equilibrium average is
\begin{eqnarray}
\langle \hat \Sigma_{x,I}(t) \rangle
&=& \hat \Sigma_x \prod_l \langle e^{-2i \hat P_{w_l(t)}} \rangle
\nonumber \\
&=& e^{-4 {\cal D}(t)} \hat \Sigma_x
\label{P17}
\end{eqnarray}
in which ${\cal D}(t)$ is given in equation (8) of the main text. Similarly, using (\ref{P12}) one obtains the dynamical correlation function
\begin{equation}
\langle \hat \Sigma_{x,I}(t) \hat \Sigma_{x,I}(t') \rangle
= e^{-4{\cal D}(t-t')} e^{4i \Phi(t,t')}
\label{P18}
\end{equation}
with phase function
\begin{eqnarray}
\Phi(t,t') &=& 4 \sum_l \frac{\Lambda_l^2}{\Omega_l^2}
\sin[\Omega_l(t-t')/2]
\nonumber \\
&&\ \ \ \ \ \ \times\ \sin(\Omega_l t/2) \sin(\Omega_l t'/2)
\nonumber \\
&=& 4 \int \frac{d\Omega}{\pi \Omega^2} {\cal J}(\Omega)
\sin[\Omega(t-t')/2]
\nonumber \\
&&\ \ \ \ \ \ \times\  \sin(\Omega t/2) \sin(\Omega t'/2)
\nonumber \\
&=& \mathrm{Im}[J(t) + J(t') - J(t-t')]
\label{P19}
\end{eqnarray}
in which we define the time-domain spectral function
\begin{eqnarray}
J(t) &=& \int \frac{d\Omega}{\pi \Omega^2} {\cal J}(\Omega) e^{-i\Omega t}
\nonumber \\
&=& \frac{2 \alpha_\Sigma \Gamma(S-1) \Omega_c^{S-1}}{(1 + i\Omega_c t)^{S-1}}.
\label{P20}
\end{eqnarray}
where the last line follows from the model form (\ref{A28}), and one therefore obtains
\begin{eqnarray}
\mathrm{Im}[J(t)] &=& -\frac{2\alpha_\Sigma \Gamma(S-1) \Omega_c^{S-1}}
{[1 + (\Omega_c t)^2]^{(S-1)/2}}
\nonumber \\
&&\times\ \sin\left[(S-1) \tan^{-1}(\Omega_c t) \right].
\label{P21}
\end{eqnarray}
Equation (\ref{P19}) is the phase function appearing in equation (9) of the main text.

\subsubsection{Application to spin--boson effective model}
\label{app:manybody}

For nonzero $H_x$, the formal solution for the reduced density matrix, generalizing equation (7) in the main text, may now be derived as follows. First, the interaction representation for the density matrix
\begin{equation}
\hat \rho(t) = e^{-i t H_{\Sigma B}} \hat \rho_0 e^{i t H_{\Sigma B}}
\label{P22}
\end{equation}
is obtained from the interaction representation
\begin{equation}
e^{it H_{\Sigma B}} = e^{it H_{\Sigma B}^0} \hat U_I(H_x t)
\label{P23}
\end{equation}
in which $H_{\Sigma B}^0 = \tilde H_B + H'_{\Sigma B}$ and
\begin{eqnarray}
\hat U_I(H_x t) &=& e^{-it H_{\Sigma B}^0} e^{it H_{\Sigma B}}
\nonumber \\
&=& T\left[e^{-i H_x \int_0^t dt' \hat \Sigma_{x,I}(t')} \right]
\nonumber \\
\hat \Sigma_I(t) &=& e^{-it H_{\Sigma B}^0} \hat \Sigma_x e^{it H_{\Sigma B}^0}
\nonumber \\
&=& \sum_\sigma e^{-it H_{B\sigma}} e^{it H_{B,-\sigma}}
\otimes |\sigma \rangle \langle -\sigma|
\nonumber \\
&=& \sum_\sigma e^{2i\sigma \hat P(t)} \otimes |\sigma \rangle \langle -\sigma|,
\label{P24}
\end{eqnarray}
where $T[\cdot]$ is the time-ordering operator (earlier times to the right), and
\begin{equation}
H_{B\sigma} = \sum_l \Omega_l \hat B_l^\dagger \hat B_l
- \sigma \sum_l \Lambda_l (\hat B_l + \hat B_l^\dagger)
\label{P25}
\end{equation}
are the spin state eigenoperators, with $\sigma = \pm 1$ in this case, obeying
\begin{equation}
H_{\Sigma B} |\sigma \rangle = H_{B\sigma} |\sigma \rangle.
\label{P26}
\end{equation}
We note that $\langle \sigma |\hat \Sigma_x| \sigma' \rangle = \delta_{\sigma,-\sigma'}$, and $\hat P(t)$ (linear in the boson operators) is defined in (\ref{P16}). The last line is equivalent to the first line of (\ref{P16}) since $\hat \Sigma_\sigma = |\sigma \rangle \langle -\sigma|$. We see that $\hat \Sigma_I(t)$ is a spin-flip operator, with amplitudes governed by the boson translation operator.

The reduced density matrix is
\begin{eqnarray}
\hat \rho_\Sigma(t) &=& \mathrm{tr}^B [\hat \rho(t)]
\nonumber \\
&=& \mathrm{tr}^B [\hat U_I(H_x t)^\dagger \hat \rho^0(t) \hat U_I(H_x t)]
\label{P27}
\end{eqnarray}
in which $\hat \rho^0(t)$ is defined by the $H_x = 0$ form
\begin{eqnarray}
\hat \rho(t) &=& \sum_{\sigma,\sigma'}
A_{\sigma \sigma'} |\sigma\rangle \langle \sigma'|
\otimes \hat \rho^B_{\sigma \sigma'}(t)
\nonumber \\
\hat \rho^B_{\sigma \sigma'}(t) &=&
e^{-it H_{B\sigma}} \hat \rho^B_\mathrm{eq} e^{it H_{B\sigma'}}.
\label{P28}
\end{eqnarray}

Defining the operator matrix elements
\begin{eqnarray}
\hat U_{I,\sigma \sigma'}(H_x t)
&=& \langle \sigma| \hat U_I(H_x t) |\sigma' \rangle
\nonumber \\
\hat U_{I,\sigma \sigma'}(H_x t)^\dagger
&=& \langle \sigma'|\hat U_I(H_x t)^\dagger |\sigma \rangle,
\label{P29}
\end{eqnarray}
one may express this in the form
\begin{equation}
\hat \rho_\Sigma(t) = \sum_{\sigma,\sigma'}
A_{\sigma \sigma'}(t) |\sigma \rangle \langle \sigma'|
\label{P30}
\end{equation}
in which, generalizing equations (7) and (8) in the main text,
\begin{eqnarray}
A_{\sigma \sigma'}(t) &=&
\mathrm{tr}^B [\langle \sigma| \hat U_I(H_x t)^\dagger
\hat \rho^0(t) \hat U_I(H_x t) |\sigma'\rangle]
\nonumber \\
&=& \sum_{\sigma_1,\sigma_2}
A_{\sigma_1 \sigma_2} M^{\sigma \sigma'}_{\sigma_1 \sigma_2}(t)
\label{P31} \\
M^{\sigma \sigma'}_{\sigma_1 \sigma_2}(t)
&=& \mathrm{tr}^B \left[\hat U_{I,\sigma_1 \sigma}(H_x t)^\dagger
\hat \rho^B_{\sigma_1 \sigma_2}(t)
\hat U_{I,\sigma_2 \sigma'}(H_x t) \right]
\nonumber \\
&=& \left\langle e^{it H_{B\sigma_2}} \hat U_{I,\sigma_2 \sigma'}(H_x t)
\hat U_{I,\sigma_1 \sigma}(H_x t)^\dagger e^{-it H_{B\sigma_1}} \right\rangle
\nonumber
\end{eqnarray}
where the $H_x = 0$ form $\hat \rho^B_{\sigma \sigma'}(t)$ is given by (\ref{P28}), the cyclic property of the trace has been used to obtain the last line, and the equilibrium average is with respect to $\hat \rho^B_\mathrm{eq}$.

Considering now small $H_x$, we note from the solution (\ref{P16}) that $\hat \Sigma_I(t)$ oscillates strongly, but remains bounded. Therefore it makes sense to expand
\begin{eqnarray}
\hat U_I(H_x t) &=& 1 - i H_x \int_0^t dt_1 \hat \Sigma_{x,I}(t_1)
\label{P32} \\
&&-\ H_x^2 \int_0^t dt_1 \int_0^{t_1} dt_2
\hat \Sigma_{x,I}(t_1) \hat \Sigma_{x,I}(t_2)
+ \ldots,
\nonumber
\end{eqnarray}
which corresponds to a sequence of $n$ spin flips at order $H_x^n$. It is apparent that $M^{\sigma \sigma'}_{\sigma_1 \sigma_2}$ corresponds to a sum of terms, starting with spin $\sigma_2$ on the left, which then undergoes a sequence of flips ending with $\sigma'$, and starting with spin $\sigma_1$ on the right, which then undergoes a sequence of flips ending with $\sigma$. The generic term in the average is of the form
\begin{widetext}
\begin{eqnarray}
M^{\sigma\sigma'}_{\sigma_1\sigma_2}(t;t_1,\ldots,t_M;t'_1,\ldots,t'_{M'})
&=& \delta_{\sigma_2, (-1)^{M'} \sigma'} \delta_{\sigma_1, (-1)^M \sigma}
\nonumber \\
&\times& \left \langle e^{it H_{B\sigma_2}}
\prod_{m'=1}^{M'} e^{-2i (-1)^{m'} \sigma_2 \hat P(t'_{m'})}
\prod_{m=1}^M e^{2i (-1)^m \sigma_1 \hat P(t_m)}
e^{-it H_{B\sigma_1}} \right \rangle
\label{P33}
\end{eqnarray}

By repeated application of the identity (\ref{A11}), the product of $\hat P$-exponentials may be combined into a single exponential of the sum, multiplied by a commutator phase factor, reducing then to the exponential of single $\hat P$ operator:
\begin{equation}
\prod_{k=1}^K e^{i \hat P_{{\bf w}_k}}
\equiv e^{i \hat P_{{\bf w}_1}} e^{i \hat P_{{\bf w}_2}}
\ldots e^{i \hat P_{{\bf w}_K}}
= e^{i \hat P_{{\bf w}^{(K)}}}
e^{-\frac{1}{2} \sum_{1 \leq k < k' \leq K}
[\hat P_{{\bf w}_k}, \hat P_{{\bf w}_{k'}}]}
\label{P34}
\end{equation}
in which
\begin{equation}
\hat P_{\bf w} = i \sum_l (w_l^* \hat B_l - w_l \hat B_l^\dagger), \ \
{\bf w}^{(K)} = \sum_{k=1}^K {\bf w}_k,\ \
[\hat P_{{\bf w}_k},
\hat P_{{\bf w}_{k'}}] = \sum_l (w_{k,l}^* w_{k',l} - w_{k,l} w_{k',l}^*).
\label{P35}
\end{equation}
Note that the indicated order of the product in (\ref{P34}) is important since the factors do not commute. A different order would result in opposite ordering of the operator pair appearing in some subset of the commutators, reversing the sign of the corresponding terms.

The final evaluation follows from the identity
\begin{eqnarray}
M({\bf u},{\bf v},{\bf w};t)
&=& \left \langle e^{it H_{B{\bf u}}} e^{2i\hat P_{\bf w}}
e^{-it H_{B{\bf v}}} \right \rangle
= \left \langle e^{i\hat P_{\bf u}}
e^{it \tilde H_B} e^{-i\hat P_{\bf u}}
e^{2i\hat P_{\bf w}} e^{i\hat P_{\bf v}}
e^{-it \tilde H_B} e^{-i\hat P_{\bf v}} \right \rangle
\nonumber \\
&=& \left \langle e^{i\hat P_{\bf u}} e^{-i\hat P_{{\bf u}_{-t}}}
e^{2i\hat P_{{\bf w}_{-t}}} e^{i\hat P_{{\bf v}_{-t}}}
e^{-i\hat P_{\bf v}} \right \rangle
= \left \langle e^{i\hat P_{{\bf u}(-t)}}
e^{2i\hat P_{{\bf w}_{-t}}} e^{-i\hat P_{{\bf v}(-t)}} \right \rangle
e^{i \sum_l (|u_l|^2 - |v_l|^2) \sin(\Omega_l t)}
\label{P36}
\end{eqnarray}
\end{widetext}
in which
\begin{eqnarray}
H_{B{\bf w}} &=& e^{i P_{\bf w}} \tilde H_B e^{-i \hat P_{\bf w}}
\nonumber \\
&=& \sum_l \Omega_l (\hat B_l^\dagger - w_l^*)(\hat B_l - w_l)
\label{P37}
\end{eqnarray}
is a generically shifted Hamiltonian, ${\bf w}_{-t} = \{w_l e^{i\Omega_l t} \}$, ${\bf w}(-t) = {\bf w} - {\bf w}_{-t} = \{w_l (1-e^{i\Omega_l t}) \}$ and the translation identity (\ref{A4}) and evolution identity (\ref{A7}) have been used [noting the change $t \to -t$ due to the cyclic reordering of the evolution operators in the last line of (\ref{P31})]. In the present application, ${\bf u} = \sigma_1 \sigma_2 {\bf v}$, hence $|u_l|^2 = |v_l|^2$, and the commutator phase factor vanishes.

The last three $\hat P$-exponentials in (\ref{P36}) may again be combined into a single one, and the final average is of the form
\begin{equation}
\langle e^{i \hat P_{\bf w}} \rangle
= e^{-\sum_l |w_l|^2 \left[n_B(\Omega_l) + \frac{1}{2}\right]},
\label{P38}
\end{equation}
in which (\ref{A20}) has been used. The explicit result is
\begin{widetext}
\begin{eqnarray}
\left \langle e^{i\hat P_{{\bf u}(-t)}}
e^{2i\hat P_{{\bf w}_{-t}}} e^{-i\hat P_{{\bf v}(-t)}} \right \rangle
&=& \left \langle e^{i\hat P_{{\bf u}(-t) + 2{\bf w}_{-t} - {\bf v}(-t)}} \right \rangle
e^{\frac{1}{2}[\hat P_{{\bf u}(-t)}, P_{{\bf v}(-t)}]
+ [P_{{\bf w}_{-t}}, \hat P_{{\bf u}(-t)}]
+ [\hat P_{{\bf w}_{-t}}, P_{{\bf v}(-t)}]}
\nonumber \\
&=& e^{-\sum_l |u_l(-t) + 2w_{l,-t} - v_l(-t)|^2
\left[n_B(\Omega_l) + \frac{1}{2}\right]}
\nonumber \\
&&\times\ e^{\sum_l \{\frac{1}{2}[u_l(-t)^* v_l(-t) - u_l(-t) v_l(-t)^*]
+ w_{l,-t}^* [u_l(-t) + v_l(-t)] - w_{l,-t} [u_l(-t)^* + v_l(-t)^*]\}}
\ \ \ \ \ \
\label{P39}
\end{eqnarray}

We finally apply this formalism to evaluate (\ref{P31}) to linear order in $H_x$:
\begin{eqnarray}
M_{\sigma_1 \sigma_2}^{(1) \sigma \sigma'}(t)
&=& -i H_x \int_0^t dt_1 \left\langle e^{it H_{B\sigma_2}}
\left[\langle \sigma_2| \hat \Sigma_I(t_1)
|\sigma' \rangle \delta_{\sigma_1 \sigma}
- \langle \sigma| \hat \Sigma_I(t_1)
|\sigma_1 \rangle \delta_{\sigma_2 \sigma'} \right]
e^{-it H_{B\sigma_1}}  \right\rangle
\nonumber \\
&=& -iH_x \sum_{\sigma_3} (\delta_{\sigma_1 \sigma}
\delta_{\sigma_2, -\sigma'} \delta_{\sigma_3 \sigma_2}
- \delta_{\sigma_1, -\sigma} \delta_{\sigma_2, \sigma'}
\delta_{\sigma_3, -\sigma_1})
\int_0^t dt_1 \left\langle e^{it H_{B\sigma_2}}
e^{2i \sigma_3 \hat P(t_1)}
e^{-it H_{B\sigma_1}} \right\rangle.
\label{P40}
\end{eqnarray}
Using (\ref{P25}) and (\ref{P16}) in (\ref{P36}), the interior thermal average is
\begin{eqnarray}
\left\langle e^{it H_{B\sigma_2}}
e^{2i \sigma_3 \hat P(t_1)}
e^{-it H_{B\sigma_1}} \right\rangle
&=& M\left[\left\{\sigma_2 \frac{\Lambda_l}{\Omega_l} \right\},
\left\{\sigma_1 \frac{\Lambda_l}{\Omega_l} \right\},
\left\{\sigma_3 \frac{\Lambda_l}{\Omega_l}(1-e^{-i\Omega_l t_1}) \right\}; t \right]
\nonumber \\
&=& e^{-\left\{\frac{1}{4} (\sigma_1-\sigma_2)^2 {\cal D}(t) + {\cal D}(t_1)
+ \sigma_3(\sigma_1 - \sigma_2) {\cal E}_2(t,t_1) \right\}}
e^{i\frac{1}{2} (\sigma_1 + \sigma_2) \sigma_3 \Phi(t,t_1)}
\label{P41}
\end{eqnarray}
in which ${\cal D}(t)$ was defined in equation (8) in the main text, and the additional real and imaginary exponents are given by
\begin{eqnarray}
{\cal E}_2(t,t_1) &=& 16 \sum_l \frac{\Lambda_l^2}{\Omega_l^2}
\left[n_B(\Omega) + \frac{1}{2} \right] \sin(\Omega_l t/2)
\sin(\Omega_l t_1/2) \cos[\Omega_l (t-t_1)/2]
\nonumber \\
&=& 16 \int \frac{d\Omega}{\pi \Omega^2} {\cal J}(\Omega)
\left[n_B(\Omega) + \frac{1}{2} \right] \sin(\Omega t/2)
\sin(\Omega t_1/2) \cos[\Omega (t-t_1)/2]
\nonumber \\
\Phi(t,t_1) &=& 16 \sum_l \frac{\Lambda_l^2}{\Omega_l^2} \sin(\Omega_l t/2)
\sin(\Omega_l t_1/2) \sin[\Omega_l (t-t_1)/2]
\nonumber \\
&=& 16 \int \frac{d\Omega}{\pi \Omega^2} {\cal J}(\Omega)
\sin(\Omega t/2) \sin(\Omega t_1/2) \sin[\Omega (t-t_1)/2]
\label{P42}
\end{eqnarray}
\end{widetext}
Substituting (\ref{P41}) and (\ref{P42}) into (\ref{P40}), one obtains the rather remarkable simplification
\begin{equation}
M_{\sigma_1 \sigma_2}^{(1) \sigma \sigma'}(t)
= \delta_{\sigma_1 \sigma_2} \delta_{\sigma, -\sigma'} H_x {\cal M}(t)
\label{P43}
\end{equation}
in which
\begin{equation}
{\cal M}(t) = 2 \int_0^t dt_1 e^{-{\cal D}(t_1)} \sin[\Phi(t,t_1)].
\label{P44}
\end{equation}
Here $M_{\sigma_1 \sigma_1}^{(1) \sigma, -\sigma}(t) = H_x {\cal M}(t)$ emerges from the case $\sigma_1 = \sigma_2$, which via the delta function prefactor in (\ref{P40}) implies $\sigma = - \sigma'$, $\sigma_3 = \sigma_1$ in the first term and $\sigma_3 = -\sigma_1$ in the second. The second distinct case $\sigma_1 = -\sigma_2$, implies $\sigma = \sigma'$ and $\sigma_3 = \sigma_2$ (in both terms). However, the resulting term
\begin{equation}
\bar {\cal M}(t) = e^{-{\cal D}(t)}
\int_0^t dt_1 e^{2 {\cal E}_2(t,t_1) - {\cal D}(t_1)}
\label{P45}
\end{equation}
is identical for the two delta function terms, leading to
\begin{equation}
M_{\sigma_1, -\sigma_1}^{(1) \sigma \sigma}(t)
= -i H_x [\bar {\cal M}(t) - \bar {\cal M}(t)] \equiv 0.
\label{P46}
\end{equation}

Similar to (\ref{A25}) (\ref{A30}), (\ref{A31}), substituting the model form (\ref{A28}) for the spectral function, one may write
\begin{equation}
\Phi(t,t_1) = 4 \alpha_\Sigma \Omega_c^{S-1}
G_{S-1}(\Omega_c t, \Omega_c t_1)
\label{P47}
\end{equation}
in which
\begin{eqnarray}
G_p(\tau,\tau_1) &=& 8 \int du u^{p-1} e^{-u}
\sin(\tau u/2) \sin(\tau_1 u/2)
\nonumber \\
&&\ \ \ \ \ \ \times\ \sin[(\tau-\tau_1)u/2]
\nonumber \\
&=& i \Gamma(p) \left\{\frac{1}{(1-i\tau)^p}
- \frac{1}{(1+i\tau)^p} \right.
\nonumber \\
&&\ \ \ \  +\ \frac{1}{(1+i\tau_1)^p} - \frac{1}{(1-i\tau_1)^p}
\label{P48} \\
&&\ \ \ \ +\ \left. \frac{1}{[1+i(\tau-\tau_1)]^p}
- \frac{1}{[1-i(\tau-\tau_1)]^p} \right\},
\nonumber
\end{eqnarray}
converging for $p > -3$ (with the apparent $\Gamma$-function singularities at $p = 0, -1, -2$ actually generating logarithmic terms).

Referencing the asymptotic forms (\ref{A32}), the thermal component of the decoherence function ${\cal D}(t)$ diverges for large $t$ in the range $0 < S < 2$ (corresponding to $-1 < p < 1$), guaranteeing convergence of the integral (\ref{P44}) for large $t$. The precise scaling of the large $t$ limit is clearly complicated, depending on the sign of $p$ and the balance between the thermal and zero temperature components of ${\cal D}(t)$, but it is clear that ${\cal M}_\infty = {\cal M}(t \to \infty)$ is generically nonzero. Due to oscillations in $\sin(\Phi)$, could accidentally vanish for certain special parameter choices, but this does not change any of the major conclusions below.

Combining the first order result (\ref{P43}) with the zeroth order result (\ref{P28}), one obtains the leading density matrix
\begin{eqnarray}
\hat \rho_\Sigma &=& \sum_\sigma
\Big\{A_{\sigma \sigma} |\sigma \rangle \langle \sigma|
\nonumber \\
&&+\ \left[H_x {\cal M}(t)
+ A_{\sigma,-\sigma} e^{-{\cal D}(t)} \right]
|\sigma \rangle \langle -\sigma| \Big\}.
\ \ \ \ \ \ \ \
\label{P49}
\end{eqnarray}
This combines equations (7)--(9) of the main text. The late time density matrix corresponds to random choice of the states
\begin{equation}
|\sigma, H_x \rangle \simeq |\sigma \rangle
+ \frac{\sigma H_x {\cal M}_\infty}{2|A_{11} - \frac{1}{2}|} |-\sigma \rangle
\label{P50}
\end{equation}
with probability $A_{\sigma \sigma}$ [$= |a_\sigma|^2$ for the pure state case $A_{\sigma \sigma'} = a_\sigma a_{\sigma'}$ with $|\Psi_\Sigma \rangle = a_+ |+\rangle + a_- |-\rangle$]. The admixture of the opposite spin state corresponds to small polarization by the transverse field. The symmetry broken state is critical here: in the absence of bath coupling term $H'_{\Sigma B}$ in equation (5) of the main text, an arbitrarily small value of $H_x$ (for $H_z = 0$) will lead to the balanced mixture $\frac{1}{\sqrt{2}}(|+ \rangle \pm |- \rangle)$. The case $A_{11} = A_{22} = \frac{1}{2}$ is singular, apparently also leading to the mixture $\frac{1}{\sqrt{2}}(|+ \rangle \pm |- \rangle)$, and requires a more careful treatment.

\subsection{Decoherence model parameter estimation}
\label{app:parmest}

We summarize here approaches to estimating the parameters in the effective model, equation (5) in the main text, from the higher level model, equation (4) in the main text.

Consider first the parameter $H_x$ that leads to tunneling between the two spin--boson model ground states $|U,\pm \rangle$.

\subsubsection{General perturbation formulation}
\label{app:genpertform}

We consider a Hamiltonian $H = H_0 + V$, in the form of a sum of unperturbed Hamiltonian $\hat H_0$ and a potential $V$. Let the system start at time $t=0$ in an initial eigenstate $|i\rangle$ of $H_0$. Then the amplitude that it will be found in (a different) eigenstate $|f \rangle$ at a later time $t$ takes the general form \cite{Merzbacher}
\begin{equation}
c_{fi}(t) = \langle f| \hat U(t,0) | i \rangle
\label{B1}
\end{equation}
in which
\begin{equation}
\hat U(t,0) = T\left[e^{-i\int_0^t dt' V_I(t')} \right]
\label{B2}
\end{equation}
is the interaction picture evolution operator, where $T$ is the time-ordering operator (earlier times to the right) and
\begin{equation}
V_I(t) = e^{it H_0} V e^{-it H_0}
\label{B3}
\end{equation}
is the interaction picture perturbation. To linear order in $V$ one obtains
\begin{equation}
c_{fi}(t) = V_{fi} \frac{1 - e^{i\omega_{fi} t}}{\omega_{fi}},
\label{B4}
\end{equation}
where
\begin{equation}
\omega_{fi} = \epsilon_f^0 - \epsilon_i^0
\label{B5}
\end{equation}
is the initial and final state energy difference, and
\begin{equation}
V_{fi} = \langle f| V | i \rangle
\label{B6}
\end{equation}
is the associated perturbation matrix element. If $V_{fi}$ vanishes, then the leading higher order contribution is the second order term
\begin{equation}
c_{fi}(t) = {\sum_n}' \frac{V_{fn} V_{ni}}{\omega_{in}}
\left(\frac{1 - e^{i\omega_{fi} t}}{\omega_{fi}}
- \frac{1 - e^{i\omega_{fn} t}}{\omega_{fn}} \right),
\label{B7}
\end{equation}
in which the sum is over the complete set of eigenstates $|n \rangle$ of $H_0$. The prime on the sum is a reminder that only terms with non-vanishing matrix elements survive. In particular, both terms $n = i,f$ are absent under the assumption $V_{fi} = V_{fi}^* = 0$.

\subsubsection{Decoherence application}
\label{app:decoherehapp}

We compare the effective model, equation (5) in the main text, identifying
\begin{equation}
H_0 = H_B + H'_{\Sigma B} - H_z \hat \Sigma_z,\ \
V = -H_x \hat \Sigma_x,
\label{B8}
\end{equation}
with the microscopic model, equation (4) in the main text, identifying
\begin{equation}
H_0 = H_{SB} + H_A,\ \
V = H^{(2)}_{AB}.
\label{B9}
\end{equation}
We do not consider the cubic interaction $H_{AB}^{(1)}$ here since it leads to the desired tunneling result only at higher order.

For the effective model, since $|\pm \rangle_\Sigma$ are defined to be eigenstates of $\hat \Sigma_z$, one obtains the finite first order result
\begin{equation}
V_{-+} = -H_x \langle -| \hat \Sigma_x |+ \rangle = -\frac{1}{2} H_x.
\label{B10}
\end{equation}
The boson sector of the unperturbed states $|\pm \rangle \otimes |B \rangle$ plays no role here.

For the microscopic model, using $|i \rangle = |+, U \rangle$, $|f \rangle = |+, U \rangle$, one obtains
\begin{equation}
\langle f| H_{AB} |i \rangle = \sum_{l,m,n} \lambda_{llmn} \bar n_{A,l}
\langle f| \hat b_m^\dagger \hat b_n |i \rangle.
\label{B11}
\end{equation}
in which
\begin{equation}
\bar n_{A,l} = \bar n(\Omega_l) = \langle \hat a_l^\dagger \hat a_l \rangle_A
\label{B12}
\end{equation}
is the mean boson count with respect to the thermal background state determined by $e^{-\beta H_A}$.

One may always write
\begin{equation}
|\sigma, U \rangle = \sum_{\sigma' = \pm} A_{\sigma \sigma'}
|\sigma' \rangle \otimes |\psi^B_{\sigma \sigma'} \rangle
\label{B13}
\end{equation}
obeying $\sum_{\sigma'}|A_{\sigma \sigma'}|^2 = 1$, in which $|\psi^B_{\sigma \sigma'} \rangle$ is the boson state obtained by projecting $|\sigma, U \rangle$ onto the $\sigma'$ spin sector.

An approximate form for these states (accurate for small $h_x$) is
\begin{equation}
|\sigma, U \rangle = \left(|\sigma \rangle - \frac{\sigma h_x}{2\Delta h_z(\alpha)}
|-\sigma \rangle \right) \otimes |\psi^B(\sigma \sqrt{\alpha}) \rangle,
\label{B14}
\end{equation}
in which
\begin{equation}
|\psi^B(\sigma \sqrt{\alpha}) \rangle
= \otimes_k |0, \sigma \sqrt{\alpha} \rangle_k
\label{B15}
\end{equation}
is the shifted boson ground state with vanishing boson occupation numbers $n_k = 0$, and
\begin{eqnarray}
\Delta h_z(\alpha) &=& \sum_k \lambda_k
\langle \psi^B(+\sqrt{\alpha})|
\hat b_k + \hat b_k^\dagger |\psi^B(+\sqrt{\alpha}) \rangle
\nonumber \\
&=& 2 \sum_k \frac{\lambda_k^2}{\omega_k}
= 2 \int \frac{d\omega}{\pi \omega} J(\omega)
\label{B16}
\end{eqnarray}
is the effective field experienced by the spin in the presence of this shifted background. To leading order in $h_x$, the spin flip within each eigenfunction occurs at fixed background boson state, with energy difference $\Delta h_z$.

Using the states (\ref{B14}), the matrix element (\ref{B11}) may be evaluated further
\begin{eqnarray}
\langle f| H_{AB} |i \rangle
&=& \sum_{l,m,n} \lambda_{llmn} \bar n_{A,l}
\langle \psi^B(-\sqrt{\alpha})| \hat b_m^\dagger
\hat b_n |\psi^B(+\sqrt{\alpha}) \rangle
\nonumber \\
&=& -M^B_{+-}(\alpha) \sum_{l,m,n} \lambda_{llmn} \bar n_{A,l}
b^0_m b^0_n,
\label{B17}
\end{eqnarray}
in which
\begin{equation}
b^0_k = \frac{1}{\sqrt{2}} q^0_k = \frac{\lambda_k}{2 \omega_k}
\label{B18}
\end{equation}
is the amplitude of the state $k$ ground state shift, and
\begin{equation}
M^B_{+-}(\alpha) = \langle \psi^B(-\sqrt{\alpha})|\psi^B(+\sqrt{\alpha}) \rangle
\label{B19}
\end{equation}
is the off-diagonal boson state overlap. The last line of (\ref{B17}) follows from the fact that $\hat b_m - b^0_m$ annihilates the shifted ground state on the right, while $\hat b_l^\dagger + b^0_n$ annihilates the oppositely shifted ground state on the left.

Although the shifted operators no longer exactly annihilate the states $|\psi^B_{\sigma \sigma'} \rangle$ in (\ref{B13}), a reasonable approximation for $h_x$ not too large is a variational approximation in which one substitutes
\begin{eqnarray}
M^B_{+-}(\alpha,h_x) &=& A_{--}^* A_{++}
\langle \psi^B_{--}|\psi^B_{++} \rangle
\nonumber \\
&&+\ A_{-+}^* A_{+-} \langle \psi^B_{-+}|\psi^B_{+-} \rangle
\label{B20}
\end{eqnarray}
in the last line of (\ref{B17}), and derived from wavefunctions still of the shifted from, but with renormalized (reduced) values for the shifts $b^0_k(h_x) < b_k^0$ obtained from the variational equations.

Comparing the effective model coefficient (\ref{B10}) with the result of the microscopic model analysis, we propose the choice
\begin{equation}
H_x = -2 \langle f|H_{AB} |i\rangle.
\label{B21}
\end{equation}
Physical estimates will be provided below.

\subsubsection{Optical model realization}
\label{app:optreal}

In the context of an optical model, $H_{AB}$ takes the real space form
\begin{equation}
H_{AB} = \frac{e^2}{2m} \int d{\bf r}
{\bf A}({\bf r})^2 \psi^\dagger({\bf r}) \psi({\bf r}),
\label{B22}
\end{equation}
in which
\begin{eqnarray}
\psi({\bf r}) &=& \frac{1}{\sqrt{V}}
\sum_{\bf k} \hat b_{\bf k} e^{i {\bf k} \cdot {\bf r}}
\label{B23} \\
{\bf A}({\bf r}) &=& \frac{1}{\sqrt{V}}
\sum_{{\bf k},\lambda} \sqrt{\frac{2\pi}{c|{\bf k}|}}
\left[\hat {\bf e}_\lambda \hat a_{\bf k} e^{i{\bf k} \cdot {\bf r}}
+ \hat {\bf e}_\lambda^* \hat a_{\bf k}^\dagger e^{-i{\bf k} \cdot {\bf r}} \right]
\nonumber
\end{eqnarray}
where $V$ is the system volume and $\hat {\bf e}_\lambda$ are orthonormal polarization vectors. This leads to
\begin{equation}
\langle f|H_{AB}|i \rangle = \frac{e^2}{2m}
M^B_{+-}(\alpha,h_x) \langle {\bf A}^2 \rangle_A
\int d{\bf r} |\psi^0({\bf r})|^2
\label{B24}
\end{equation}
in which translation invariance of the photon illumination is assumed so that the (thermal) average
\begin{equation}
\langle {\bf A}({\bf r})^2 \rangle_A
= \int \frac{d{\bf k}}{\pi^2 ck} \left[n_B(ck) + \frac{1}{2} \right]
\label{B25}
\end{equation}
factors out of the integral, and $\psi^0({\bf x}) = \sum_{\bf k} b_{\bf k}^0 e^{i{\bf k} \cdot {\bf r}}$ is the real space form of the ground state shift. The coefficient $e^2$ should be viewed here as an appropriate effective charge-related coupling constant.

Note that for the problem considered here, the initial and final states are degenerate in energy, $\omega_{fi} = 0$ (assuming $h_z = 0$), and (\ref{B4}) one obtains the linear transition rate
\begin{equation}
c_{fi}(t) = -i V_{fi} t,
\label{B26}
\end{equation}
valid so long as $|c_{fi}(t)| \ll 1$, beyond which the more complicated physics of the entire decoherence process takes over, nominally described by the effective model, equation (5) in the main text. In essence, $c_{fi}(t)$ represents scattering of environmental fluctuations between the two states $|\pm, U \rangle$, rendering them visible to each other.

There are other forms one could choose for $H_{AB}$ but whose linear matrix elements (\ref{B6}) vanish, necessitating more complicated analyses of the second order form (\ref{B7}).

\subsubsection{Physical estimates}
\label{app:physest}

It is clear from the form (\ref{B23}) that the tunneling rate is a product of an (optical) illumination intensity, the (mesoscopic) integrated symmetry breaking amplitude, and the matrix element describing the overlap of the symmetry broken boson states.

For shifted ground states of the form
\begin{eqnarray}
\Psi_\pm &=& \prod_k \psi_{k,0}(q_k \pm q_k^0)
\nonumber \\
\psi_{k,0}(q) &=& \left(\frac{\omega_k}{\pi} \right)^{1/4} e^{-\omega_k q^2/2}
\label{B27}
\end{eqnarray}
written in coordinate representation, one obtains
\begin{eqnarray}
\langle \Psi_- | \Psi_+ \rangle &=& \prod_k \int dq_k
\psi_{k,0}(q_k - q_k^0) \psi_{k,0}(q_k + q_k^0)
\nonumber \\
&=& e^{-\sum_k \omega_k (q_k^0)^2}
\nonumber \\
&=& e^{-\int d\omega J(\omega)/2\pi \omega}
\nonumber \\
&=& e^{-\alpha \Gamma(s) \omega_c^s},
\label{B28}
\end{eqnarray}
in which the final result follows from the form (\ref{B18}) and the spectral function, equation (2) in the main text. More generally, for finite $h_x$ one may expect to be able to write $q_k^0 = \lambda_k/2 \tilde \omega_k(h_x)$ and
\begin{equation}
\sum_k (q_k^0)^2 \delta(\omega - \omega_k) = \frac{\pi J(\omega)}{4\tilde \omega(\omega,h_x)^2}
\label{B29}
\end{equation}
for some smoothly renormalized frequency function $\tilde \omega(\omega,h_x)$. One then obtains (\ref{B27}) with $\omega$ replaced by $\tilde \omega^2/\omega$ in the integrand denominator. The essential conclusion is that the matrix element may be expected to be $O(1)$, except for perhaps extreme choices of $\alpha \omega_c^s$ where the shifted ground state overlaps become exponentially small.

In order to provide a physical estimate for (\ref{B21}), one sees that the combination $\frac{1}{2} e^2 \langle {\bf A}^2 \rangle$ acts like a chemical potential coupled to a particle number $\int d{\bf r} |\psi^0|^2$ measuring the total boson shift accompanying the magnetic symmetry breaking $\langle \hat S_z \rangle$.

Thermal photon number per unit volume (reinserting physical units):
\begin{eqnarray}
\bar n_\mathrm{ph} &=& \int \frac{d{\bf k}}{(2\pi)^3} n_B(\hbar c k)
\nonumber \\
&=& \left(\frac{k_B T}{2\pi \hbar c}\right)^3 \int_0^\infty \frac{u^2 du}{e^u - 1}
\nonumber \\
&=& \Gamma(3)\zeta(3) \left(\frac{k_B T}{2\pi \hbar c}\right)^3.
\label{B30}
\end{eqnarray}
Corresponding average of the thermal part of ${\bf A}^2$:
\begin{eqnarray}
\frac{e^2}{2m} \langle {\bf A}^2 \rangle_\mathrm{A,th}
&=& \frac{4\pi e^2 \hbar^2}{m} \int \frac{d{\bf k}}{(2\pi)^3}
\frac{n_B(\hbar ck)}{\hbar ck}
\nonumber \\
&=& \frac{e^2 (k_B T)^2}{2 \pi^2 \hbar m c^3}
\int_0^\infty \frac{udu}{e^u-1}
\nonumber \\
&=& \Gamma(2) \zeta(2) \frac{e^2}{\hbar c}
\frac{(k_B T)^2}{2\pi^2 m c^2}.
\label{B31}
\end{eqnarray}
Integral of symmetry broken state defined by number density $\rho^0({\bf r}) = |\psi^0({\bf r})|^2$:
\begin{equation}
N^0 = \int_V d{\bf r} \rho^0({\bf r}) \equiv \bar \rho^0 V^0,
\label{B32}
\end{equation}
where $V^0$ defines an effective macroscopic support volume of the broken symmetry region. The transition rate is therefore estimated as
\begin{equation}
\frac{1}{\hbar} \langle f| H_{AB} |i \rangle
= e^{-\alpha \Gamma(s) \omega_c^s}
\frac{\Gamma(2) \zeta(2)}{2\pi^2} \frac{e^2}{\hbar c}
\frac{k_B T}{m c^2} \frac{k_B T}{\hbar} N^0.
\label{B33}
\end{equation}
The factor $k_BT/mc^2$ is a thermal-relativistic energy ratio and $k_BT/\hbar$ is a characteristic thermal frequency.

The fine structure constant is $e^2/\hbar c = 1/137$. Let $V^0 = 1$ mm$^3$ be a pixel volume, and let $(\rho^0/\rho_e)^{1/3} = 10^{-3}$ correspond to a 0.1\% linear shift in the electron cloud with $m = m_e = 9.11 \times 10^{-31}$ being the electron mass. At room temperature $T = 293$ K one obtains $k_BT/m_ec^2 = 4.95 \times 10^{-8}$ and $k_BT/\hbar = 3.84 \times 10^{13}$ 1/s. Let the electron number density be estimated (from Si) as $\rho_e = 14 N_A (2.33$ g/cm$^3)/(28.1$ g/mol$) = 6.99 \times 10^{23}$ electrons/cm$^3$, where $N_A = 6.02 \times 10^{23}$ is Avogadro's number. Thus, $N^0 = 10^{-9} \rho_e V^0 = 6.99 \times 10^{11}$ electrons. Using the zeta function value $\zeta(2) = 1.6449$, these combine to yield
\begin{equation}
\frac{\Gamma(2) \zeta(2)}{2\pi^2} \frac{e^2}{\hbar c}
\frac{k_B T}{m c^2} \frac{k_B T}{\hbar} N^0
= 8.07 \times 10^{14} \ 1/\mathrm{s}.
\label{B34}
\end{equation}
This is an enormous rate, and will remain enormous for any reasonable choice of the matrix element exponential factor. If one uses instead the cryogenic temperature $T = 1$ K, the result (\ref{B34}) reduces to $10^{10}$ 1/s, which is still extremely large. This is the estimate for $H_x/\hbar$ quoted in the paragraph below equation (5) in the main text.


\begin{thebibliography}{}
\bibitem{foot:T0decoh} At $T=0$ the $n_B(\Omega)$ term is dropped in (\ref{9}). The resulting ground state decoherence result ${\cal D}_0(t)$ essentially replaces $S \to S+1$ in the finite $T$ result. In particular, full decoherence occurs for $-1 \leq S \leq 1$, with ${\cal D}_0(t) \approx -4\alpha_\Sigma \Gamma(S-1) \cos[\pi(S-1)/2] t^{1-S}$, while for $S > 1$ one obtains the finite limit ${\cal D}_0(t) \to 4\alpha_\Sigma \Gamma(S-1) \Omega_c^{S-1}$.
\end{thebibliography}
\begin{thebibliography}{}
\bibitem{foot:rho} As is standard, since the density matrix is built out of Schr\"odinger picture states evolving forward in time, it evolves in reverse time compared to Heisenberg picture operators: $\hat A(t) = e^{it H} \hat A(0) e^{-it H}$.
\end{thebibliography}
\end{document}